\documentclass[twocolumn,aps,showpacs,preprintnumbers,amsmath,amssymb]{revtex4}
\usepackage{graphicx}

\begin{document}

\title{Strong subadditivity inequality for quantum entropies and four-particle entanglement}

\author{Asoka Biswas and G. S. Agarwal}
\affiliation{Physical Research Laboratory, Navrangpura, Ahmedabad - 380 009, India}
\date{\today}

\begin{abstract}
Strong subadditivity inequality for a three-particle composite system is an 
important inequality in quantum information theory which can be studied via a four-particle 
entangled state. We use two three-level atoms in $\Lambda$ configuration interacting with a 
two-mode cavity and the Raman adiabatic passage technique for the production of the 
four-particle entangled state. Using this four-particle
entanglement, we study for the first time various aspects of the strong subadditivity inequality. 
\end{abstract}


\maketitle

\section{\label{sec:intro}Introduction}

Entanglement \cite{schro} in a composite system refers to certain implicit correlation 
between the subsystems arising from their interaction. It is the key resource
of quantum computation and quantum information processing \cite{chuang}. Due to recent 
advances in this field, entanglement has generated renewed interest. There have 
been different approaches to understand and to quantify entanglement \cite{liter}. But so 
far the entanglement, only in a bipartite pure state has been investigated 
very extensively. The von Neumann entropy \cite{neumann} of either of the subsystems provides a 
good measure of entanglement in this case \cite{von}. This is the quantum 
partner of the Shannon's entropy \cite{shannon} in classical information theory and is defined as \cite{wehrl}
\begin{equation}
\label{entropy}S(j)=-\textrm{Tr}_j(\rho_j\log_2\rho_j)\;,
\end{equation}
where $j=A, B$. Here, $\rho_j$ is the reduced density 
operator of the subsystem $j$ and is given by 
\begin{equation}
\rho_j=\textrm{Tr}_l\rho_{AB}\;,
\end{equation}
where $\rho_{AB}$ is the density operator of the composite system under consideration and 
$j,l=A, B$, $j\neq l$.
In general, the quantities $S(j)$ satisfy the following inequality (due to Araki and Lieb) \cite{araki}:
\begin{equation}
\label{ineq}|S(A)-S(B)|\leq S(A,B)\leq S(A)+S(B)\;,
\end{equation}
where $S(A,B)$ is the joint entropy of the composite system comprising $A$ and $B$. 
The second part of the above inequality is known as subadditivity inequality \cite{ruskai}. For 
a pure state, $S(A,B)=0$ and thus $S(A)=S(B)$. The equality sign in the above relation 
holds good if and only if the composite density matrix $\rho_{AB}$ can be
written as a tensor product of its two reduced density matrices $\rho_A$ and $\rho_B$, i.e., for a disentangled state. 
One can define 
the index of correlation $I_c$ given by the expression $S(A)+S(B)-S(A,B)$ \cite{phoenix}, 
which can also be interpreted as information entropy in quantum information point of view. 
We note that Kim {\it et al.\/} have calculated the entropies of 
different kinds of pure states including two-mode Fock states and squeezed states \cite{kim}.
Further, the above relation for entropy has been studied in the context of entangled 
Gaussian states \cite{gsa}.

So far we have discussed about the measurement of entanglement in a bipartite pure state. 
If the composite system is in a mixed state (defined by the density operator $\rho$), 
the entanglement of formation $E_F$ can be defined in terms of the average von Neumann 
entropies of the pure states of the decompositions \cite
{von}. Wootters has shown the quantity $E_F$ to be an explicit function of $\rho$ \cite{woot}. 
He has introduced the notion of concurrence in this context.  

We further notice that from the Schmidt decomposition of a pure bipartite state,
one can properly identify the entanglement present in the state \cite{peres}. 
This is also very useful to study bipartite continuous systems \cite{law}. 
On the other hand, for a mixed state $\rho$, the separability criterion has 
been proposed \cite{criterion} to study entanglement. This is based on positive partial 
transpose mapping of $\rho$. Thus 
the negativity (entanglement monotone) of the eigenvalues of the partial 
transpose of $\rho$ could be a measure of entanglement in a mixed 
bipartite system \cite{vidal}. The concept of negativity as an entanglement measure 
has been used in context of interaction of atoms with thermal field \cite{kim1}.
The separability criterion has been extended to continuous systems \cite{simon} also. 

Despite many approaches to define entanglement for a bipartite system, there 
have been only a few approaches to quantify entanglement in the composite 
systems of three or more particles \cite{woot,vedral,thapliyal}. We note that a generalization of Schmidt 
decomposition in multipartite systems in pure states has been introduced \cite{cart}. Coffman {\it et al.\/} 
\cite{coffman} proposed a 
measurement of entanglement in a tripartite system in terms of concurrences of 
the pairs of subsystems. This measure is invariant under permutations of the subsystems. 
An average entanglement in a four-partite entangled state has been 
defined in terms of von Neumann entropies  of the pairs of 
subsystems \cite{higuchi}. Very recently, 
Yukalov has addressed the question more generally and quantified 
multipartite entanglement \cite{yukalov} in terms of the ratio of norms of an entangling operator and of 
a disentangling operator in the relevant disentangled Hilbert space.  

In this paper, we put forward a possible measurement of entanglement of a four-particle system 
by studying the entropy of the reduced three-particle system. 
As mentioned above, the von Neumann entropy is a good measure for entanglement in a bipartite
system.
For a tripartite composite state, this entropy satisfies a strong subadditivity inequality (SSI) \cite{chuang}, which has 
many important implications in the subject of quantum information theory. In this paper,
we study the properties of a four-particle entangled state through the three-particle entropy and the SSI.

The structure of the paper is as follows. In Sec.~II, we provide 
a brief discussion on strong subadditivity inequality from the quantum information 
point of view. In Sec.~III, we describe a physical model and show the preparation of 
a four-particle entangled state. In Sec.~IV, we
study the validity of the SSI in the present context and provide a physical explanation of the 
results. We conclude this paper by proposing a measurement of the corresponding four-particle entanglement.

\section{Strong subadditivity inequality}
We have already mentioned that for a bipartite composite system of two particles $A$ and $B$, 
the joint entropy $S(A,B)$ satisfies the subadditivity inequality (\ref{ineq}). 
For a composite system of three particles A, B, and C, this inequality can be 
extended to the following form \cite{inequal}:
\begin{equation}
S(A, B, C) + S(B) \leq S(A, B) + S(B, C)\;.
\label{lieb}
\end{equation}
This inequality is known as strong subadditivity inequality. 
The most obvious situation that the equality sign holds in (\ref{lieb}) is when
the composite density matrix $\rho_{ABC}$ can be written as the tensor product
of its three reduced density matrices as $\rho_A\otimes\rho_B\otimes\rho_C$, i.e., when the system is in a disentangled state.
However, the more stringent condition for this reads as \cite{ruskai} 
\begin{equation}
\label{must}\log_2(\rho_{ABC})-\log_2(\rho_{AB})=\log_2(\rho_{BC})-\log_2(\rho_B)\;.
\end{equation}

There have been numerous implications of the above inequality (\ref{lieb}) in 
quantum information theory \cite{chuang}. Firstly, it refers to the fact that the conditioning
on the subsystem always reduces the entropy, i.e., $S(A | B, C)\leq S(A | B)$, where, 
$S(A |B) = S(A,B)-S(B)$ is the entropy of A conditional on knowing the state of B. 
Secondly, the above inequality implies that discarding a quantum system never increases 
mutual information, i.e., $S(A:B)\leq S(A:B,C)$, where, $S(A:B)=S(A)+S(B)-S(A,B)$ is the
mutual information of the subsystems A and B. Thirdly, quantum operations never 
increase mutual information of two subsystems. This means that if the mutual information 
of the two subsystems A and B becomes $S'(A:B)$ after trace-preserving 
operation on B, then $S'(A:B)\leq S(A:B)$. Further, this inequality (\ref{lieb})
implies that the conditional entropy of 
the subsystems A, B, and C is also subadditive, i.e., $S(A, B|C)\leq S(A|C)+S(B|C)$.

To verify SSI, one needs to calculate the entropies like $S(A, B, C)$ which 
clearly requires a three-particle mixed state which we can 
produce using a pure four-particle entangled state \cite{chuang1}. In the next section, 
we discuss how one can prepare a pure four-particle entangled state so that we can study 
SSI for the first time for a system realizable using cavity QED methods.

\begin{figure*}
\scalebox {0.45}{\includegraphics{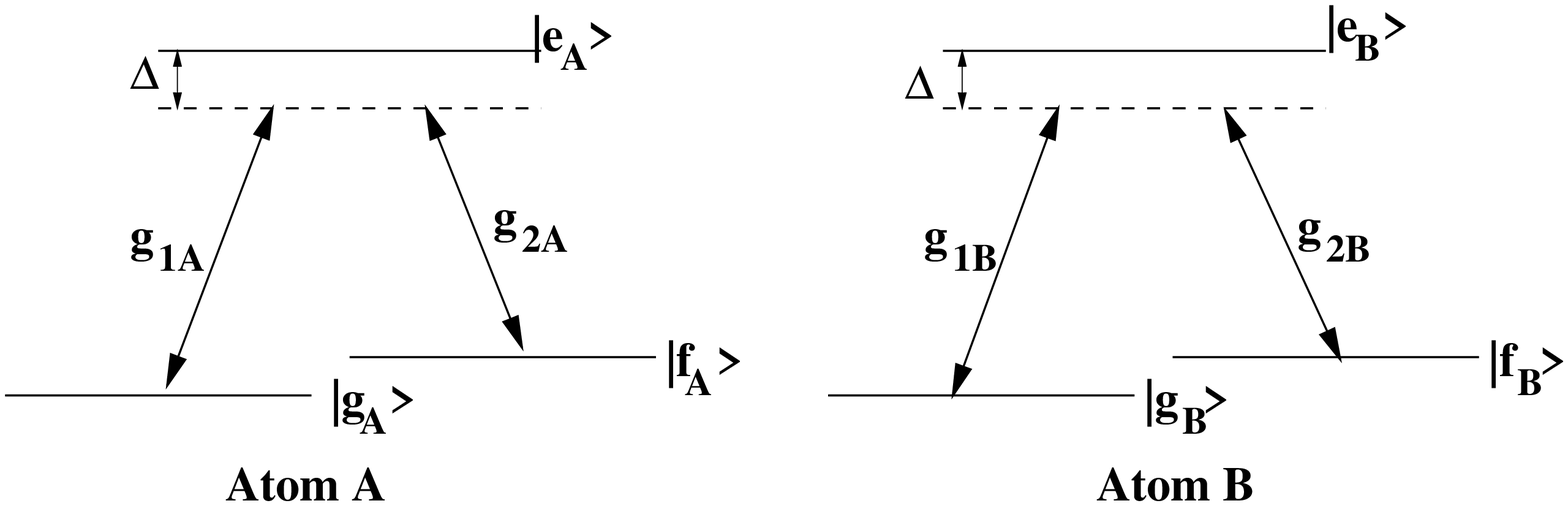}}
\caption{\label{fig1}Level diagram  of two three-level atoms in
$\Lambda$-configuration, interacting with two cavity modes defined by annihilation operators
$a$ and $b$. $g_{1k}$ and $g_{2k}$ $(k=$ A,B)
are the atom-cavity coupling terms for the $k$-th atom. $\Delta$ is the common one-photon
detuning of the fields.}
\end{figure*}

\section{\label{sec:basic}Preparation of four-particle entangled state}
We consider two three-level atoms (A and B) with relevant energy levels in
$\Lambda$-configuration (see Fig.~\ref{fig1}) interacting with a two-mode high quality 
optical cavity. The specified annihilation operators for the 
cavity modes are $a$ and $b$. The atoms are interacting with the cavity mode $a$ in 
$|e\rangle \leftrightarrow |g\rangle$ transition and with the mode $b$ in 
$|e\rangle \leftrightarrow |f\rangle$ transition. 

The Hamiltonian for the system under rotating wave approximation can be written as 
\begin{widetext}
\begin{equation}
\label{hamil1}H=\hbar(\omega_1 a^\dag a+\omega_2 b^\dag b)+\hbar\sum_{k=A,B}\left[\omega_{e_kg_k}|e_k\rangle\langle e_k|+\omega_{f_kg_k}|f_k\rangle\langle f_k|+\left\{g_{1k}|e_k\rangle\langle g_k|a+g_{2k}|e_k\rangle\langle f_k|b+\textrm{h.c.}\right\}\right]\;,
\end{equation}
where, $\omega_{l_km_k}$ is the atomic transition frequency 
between the levels $|l_k\rangle$ and $|m_k\rangle$, 
$\omega_j$ $(j=1,2)$ is the respective frequency of the cavity modes $a$ and 
$b$, $g_{jk}$ $(j=1,2)$ provides the atom-cavity coupling. We assume $g_{jk}$'s to be real and function of time.

We start with the initial state $|\psi_i\rangle =|g_A,g_B,n,\mu\rangle$, where
$n$ and $\mu$ are the initial numbers of photons in the
cavity modes $a$ and $b$, respectively and the two atoms are in $|g\rangle$ 
state.
The state of the system can be expanded in terms of the relevant basis 
states in the following way: 
\begin{eqnarray}
|\psi(t)\rangle &=& c_1|g_A,g_B,n,\mu\rangle +c_2|g_A,e_B,n-1,\mu\rangle +c_3|g_A,f_B,n-1,\mu +1\rangle\nonumber\\
&& +c_4|e_A,f_B,n-2,\mu +1\rangle +c_5|e_A,e_B,n-2,\mu\rangle +c_6|f_A,f_B,n-2,\mu +2\rangle \nonumber \\
&& +c_7|f_A,e_B,n-2,\mu +1\rangle +c_8|f_A,g_B,n-1,\mu +1\rangle +c_9|e_A,g_B,n-1,\mu\rangle \;.\label{wavefunc}
\end{eqnarray}
From the Schr{\"o}dinger equation we find the following equations of the 
corresponding probability amplitudes:
\begin{eqnarray}
\dot{d}_1&=&-i(\sqrt{n}g_{1B}d_2+\sqrt{n}g_{1A}d_9)\;,\nonumber\\
\dot{d}_2&=&-i(\sqrt{n}g_{1B}d_1+\Delta d_2+\sqrt{\mu +1}g_{2B}d_3+\sqrt{n-1}g_{1A}d_5)\;,\nonumber\\
\dot{d}_3&=&-i(\sqrt{\mu +1}g_{2B}d_2+\sqrt{n-1}g_{1A}d_4)\;,\nonumber\\
\dot{d}_4&=&-i(\sqrt{n-1}g_{1A}d_3+\Delta d_4+\sqrt{\mu +1}g_{2B}d_5+\sqrt{\mu +2}g_{2A}d _6)\;,\nonumber\\
\label{doteq}\dot{d}_5&=&-i[\sqrt{n-1}(g_{1A}d_2+g_{1B}d_9)+\sqrt{\mu +1}(g_{2B}d_4+g_{2A}d_7)+2\Delta d_5]\;,\\
\dot{d}_6&=&-i(\sqrt{\mu +2}g_{2A}d_4+\sqrt{\mu +2}g_{2B}d_7)\;,\nonumber\\
\dot{d}_7&=&-i(\sqrt{\mu +1}g_{2A}d_5+\sqrt{\mu +2}g_{2B}d_6+\Delta d_7+\sqrt{n-1}g_{1B}d _8)\;,\nonumber\\
\dot{d}_8&=&-i(\sqrt{n-1}g_{1B}d_7+\sqrt{\mu +1}g_{2A}d_9)\;,\nonumber\\
\dot{d}_9&=&-i(\sqrt{n}g_{1A}d_1+\sqrt{n-1}g_{1B}d_5+\sqrt{\mu +1}g_{2A}d_8+\Delta d_9)\;,\nonumber
\end{eqnarray}
where, we have used the following transformations:
\begin{eqnarray}
&&c_1=d_1\;,~~c_2e^{-i\Delta t}=d_2\;,~~c_3=d_3\;,~~c_4e^{-i\Delta t}=d_4\;,~~c_5e^{-2i\Delta t}=d_5\;,\nonumber\\
&&c_6=d_6\;,~~c_7e^{-i\Delta t}=d_7\;,~~c_8=d_8\;,~~c_9e^{-i\Delta t}=d_9,~~\Delta_k=\omega_{e_kl_k}-\omega_{1,2}\;,
\end{eqnarray}
where, $l_k=g_k, f_k$, $\Delta_k$ is the one-photon detuning of the cavity modes for the $k$-th atom. Here we have
assumed that the cavity modes are in two-photon resonance and  
$\Delta_A=\Delta_B=\Delta$.

Writing these equations (\ref{doteq}) in the matrix form $[\dot{d}_i]=-i[M][d_i]$, 
we find that one of the eigenvalues of the matrix $[M]$ is zero. The corresponding eigenstate is 
\begin{equation}
|\psi_0\rangle = \frac{1}{P}\left[\alpha|g_A,g_B,n,\mu\rangle +\beta|f_A,f_B,n-2,\mu +2\rangle-\gamma|g_A,f_B,n-1,\mu +1\rangle-\delta|f_A,g_B,n-1,\mu +1\rangle)\right]\;,
\label{dark}
\end{equation}
\end{widetext}
where
\begin{eqnarray}
\alpha&=&g_{2A}g_{2B}\sqrt{(\mu +1)(\mu +2)}\;,\;\beta=g_{1A}g_{1B}\sqrt{n(n-1)}\;,\nonumber\\
\label{norm}\gamma&=&g_{1B}g_{2A}\sqrt{n(\mu +2)}\;,\;\delta=g_{1A}g_{2B}\sqrt{n(\mu +2)}\;,\\
P&=&\sqrt{\alpha^2+\beta^2+\gamma^2+\delta^2}\;.\nonumber
\end{eqnarray}
Clearly, this state is an entangled state of four particles, namely, the atoms 
A and B, and the two modes $a$ and $b$. Using appropriate time-dependence of
the pulses, the four-particle system can be prepared in this state, as discussed 
in the next section. Recently, there have a few 
experimental demonstrations of preparation of four-particle entangled state 
\cite{wineland} and performance of a
C-NOT gate \cite{wineland1}. Interestingly, the state $|\psi_0\rangle$ is  a
two-atom two-mode multipartite coherent population trapping (CPT) state 
which is a counterpart of the well-known CPT state for a single atom 
interacting with two coherent fields \cite{arim}.

\section{Study of strong subadditivity inequality}
We next discuss how the state $|\psi_0\rangle$  can be prepared by using 
Raman adiabatic passage technique.
We assume that both the atoms are initially in $|g\rangle$ state. We further assume the 
time-dependence of the Rabi frequencies $g_{jk}$ of the two modes as
\begin{eqnarray}
g_{1A}=g_{1B}&=&g_{10}\exp{(-(t-T)^2/\tau^2)}\;;\nonumber\\
\label{pulse}g_{2A}=g_{2B}&=&g_{20}\exp{(-t^2/\tau^2)}\;.
\end{eqnarray}
Here, $g_{j0}$ ($j=1,2$) is the amplitude of the respective pulse, $\tau$ 
and $T$ are the width and time-separation respectively, of the two pulses.
Note that the pulses are applied in counterintuitive sequence. Under this 
condition, the atom-cavity system follows the evolution of the state 
$|\psi_0\rangle$  adiabatically.
This state is a zero eigenvalue eigenstate (adiabatic state) of the 
Hamiltonian (\ref{hamil1}).  During this process, known as stimulated Raman 
adiabatic technique (STIRAP) \cite{stirap}, the atom-cavity system remains 
in this state for all times. In the present case, at the end of the evolution, 
the population of both the atoms are simultaneously transferred to the state 
$|f\rangle$. However, if the atoms are not in one-photon resonance, i.e., if 
$\Delta\tau\neq 0$, then this transfer process is not complete. This happens because
the system does not remain confined in the null adiabatic state 
$|\psi_0\rangle$  for $\Delta\tau\neq 0$ \cite{stirap}. 

\begin{figure}
\scalebox{0.4}{\includegraphics{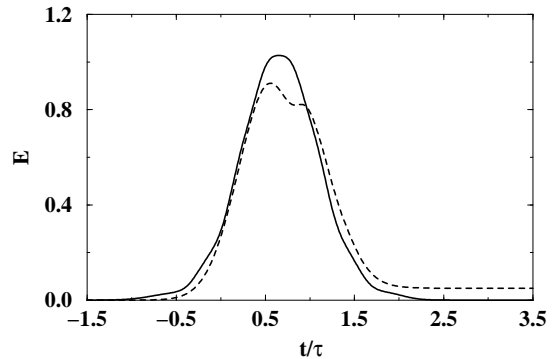}}
\caption{\label{fig2}Time evolution of the parameter $E$ for $\Delta\tau=0$ (solid curve)
and $\Delta\tau=60$ (dashed curve). This clearly shows that the strong subadditivity inequality remains 
valid in the present physical situation. The parameters chosen here are 
$n=2$, $\mu=0$, $g_{j0}\tau=15~ (j=1,2)$, $T=4\tau/3$.}
\end{figure}

We now investigate the validity of SSI for any trio of quantum systems in the present process. 
We can express this inequality for any three 
particles, namely, atom A, atom B, and cavity mode $a$ with number $n$ of photons
out of the four-particle system under consideration as 
\begin{equation}
\label{ssi}E=S(A,B)+S(A,n)-S(A,B,n)-S(A)\geq 0\;.
\end{equation}
Here, $S$ defines the joint von Neumann entropy of the relevant subsystems 
[see Eq.~(\ref{entropy})]. This can be calculated from the 
state (\ref{wavefunc}) by tracing over the other subsystems, e.g.,
\begin{equation}
S(A,B)=-\textrm{Tr}_{AB}(\rho_{AB}\log_2\rho_{AB})\;,
\end{equation}
where, $\rho_{AB}$ is the reduced density matrix of the atoms A and B and is given by 
\begin{equation}
\rho_{AB}=\textrm{Tr}_{n,\mu}(|\psi (t)\rangle\langle \psi (t)|)\;.
\end{equation}
We show the time variation of $E$ in Fig.~\ref{fig2}. Clearly, $E(t)$ never 
becomes negative during the evolution and thus the SSI (\ref{ssi}) holds for all times. 

\begin{figure}
\scalebox{0.35}{\includegraphics{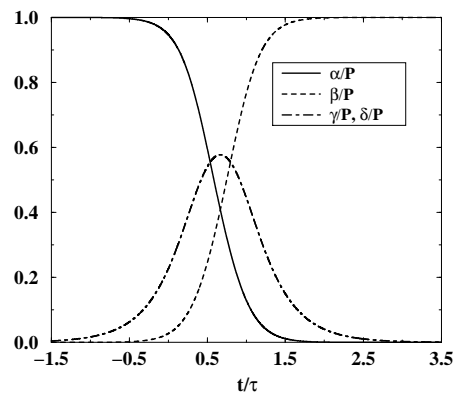}}
\caption{\label{fig3}Variation of the coefficients $\alpha/P$, $\beta/P$, $\gamma/P$, 
and $\delta/P$ with time. The parameters chosen here are $n=2$, $\mu=0$, 
$g_{j0}\tau=15$, $T=4\tau/3$, and $\Delta\tau=0$.}
\end{figure}

From Fig.~\ref{fig2}, one clearly sees that for $\Delta\tau=0$, in long time
limit, $E$ becomes zero. This means that the subsystems (A, B, and the mode $a$ with photon
number $n$) become disentangled. This happens because of complete adiabatic transfer 
of population to the level $|f\rangle$ of both the atoms at long time limit. The entire process
can be written as
\begin{equation}
\label{evolve}|g_A,g_B,n,\mu\rangle\longrightarrow|f_A,f_B,n-2,\mu+2\rangle\;.
\end{equation} 
We have shown the time-variation of the coefficients $\alpha/P$, $\beta/P$, 
$\gamma/P$, and $\delta/P$ [see Eq.~(\ref{norm})] in Fig.~\ref{fig3}.
This figure reveals the above evolution according to the state $|\psi_0\rangle$
under the action of the pulses (\ref{pulse}).
But for $\Delta\tau=60$, since complete population transfer does not occur, the
system remains entangled in the state $|\psi\rangle$ at long time limit. 
This is clear from the dashed curve of Fig.~\ref{fig2}, as
the equality $E=0$ no longer holds at this time limit.

Thus we can recognize the expression $E$ [see Eq.~(\ref{ssi})] as a measure of 
four-particle entanglement in the present process.
Precisely, $E\geq 0$, where the equality sign holds good for the disentangled states. An increase in value of $E$ refers to 
increase in entanglement. Thus, during the evolution, the system gets more entangled for $\Delta\tau=0$ than for 
$\Delta\tau=60$. However, at the end of the evolution, the entanglement persists for 
nonzero $\Delta\tau$. We must emphasize here that, the present definition of entanglement 
measurement satisfies all the relevant
criteria, namely, (a) it is semipositive, i.e., $E\geq 0$, (b) $E=0$ for an 
disentangled state, and (c) the function $E(t)$ is continuous in time domain. 

Using our four-particle entanglement, we can also study the inequality (\ref{ineq}) 
involving the entropies of two particles, say, atoms A and B.  
They remain strongly correlated during the 
evolution, as $I_c$ remains much larger than zero. At the end of the evolution, 
$I_c$ becomes zero for $\Delta\tau =0$ which means that the subsystems become 
uncorrelated. In other words, the entanglement between them vanishes.

We note in passing that for a four-particle GHZ state defined by
\begin{equation}
|\psi\rangle =\frac{1}{\sqrt{2}}(|0000\rangle+|1111\rangle)
\end{equation}
of four qubits A, B, C, and D, one is led to a three-particle mixed state
$\rho_{ABC}$ defined by
\begin{equation}
\label{mix}\rho_{ABC}=\frac{1}{2}(|000\rangle\langle 000|+|111\rangle\langle 111|)\;.
\end{equation}
In this case, $S(A, B, C)=S(A)=S(A,B)=S(B,C)=\log_22=1$. Therefore, the parameter
$E$ in this case becomes zero, as from Eq.~(\ref{ssi}). So we have a counter-example, in which 
the equality sign in (\ref{lieb}) holds for an entangled state, too. However, we
note that the above state (\ref{mix}) satisfies the condition (\ref{must}) and 
thus the said equality.

\section{conclusions}
In conclusion, we have shown for the first time the role of strong subadditivity
inequality for entropies in a four-particle composite system. The stimulated Raman
adiabatic technique has been used to prepare the four-particle entangled state 
using two three-level atoms initially in their ground states in a two-mode 
cavity.   
We further show that the parameter $E$ could serve as a possible measurement of entanglement in the four-particle entangled state under consideration.

\end{document}